\documentclass[twocolumn,aps,prb,amssymb,amsmath,floatfix,showpacs]{revtex4}
\usepackage{graphicx,subfigure}
\usepackage{bm}
\usepackage{verbatim}
\usepackage{amsmath}
\usepackage{amssymb}
\usepackage[T1]{fontenc}
\usepackage{ae,aecompl}

\newcommand{\rv}{{\bf r}}

\newcommand{\be}{\begin{equation}}
\newcommand{\ee}{\end{equation}}
\newcommand{\beq}{\begin{eqnarray}}
\newcommand{\eeq}{\end{eqnarray}}
\newcommand{\bse}{\begin{subequations}}
\newcommand{\ese}{\end{subequations}}
\newcommand{\beqq}{\begin{eqnarray*}}
\newcommand{\eeqq}{\end{eqnarray*}}

\def\rf#1{(\ref{#1})}

\begin{document}

\title{Quantum Indistinguishability in Chemical Reactions}

\author{Matthew P. A. Fisher}
\affiliation{Department of Physics, University of California, Santa Barbara,
CA 93106, USA}
\email{mpaf@kitp.ucsb.edu}
\author{Leo Radzihovsky} 
\affiliation{Center for Theory of Quantum Matter, Department of
  Physics, University of Colorado, Boulder, CO 80309, USA,}
\affiliation{Kavli Institute for Theoretical Physics, University of
  California, Santa Barbara, CA 93106, USA}
\email{radzihov@colorado.edu}
\date{\today}

\begin{abstract}
  Quantum indistinguishability plays a crucial role in many low-energy
  physical phenomena, from quantum fluids to molecular
  spectroscopy. It is, however, typically ignored in most high
  temperature processes, particularly for ionic coordinates,
  implicitly assumed to be distinguishable, incoherent and thus
  well-approximated classically.  We explore chemical reactions
  involving small symmetric molecules, and argue that in many
  situations a full quantum treatment of collective nuclear degrees of
  freedom is essential. Supported by several physical arguments, we
  conjecture a ``Quantum Dynamical Selection'' (QDS) rule for small
  symmetric molecules that precludes chemical processes that involve
  direct transitions from orbitally non-symmetric molecular states.
  As we propose and discuss, the implications of the
  Quantum Dynamical Selection rule include: (i) a differential
  chemical reactivity of para- and ortho-hydrogen, (ii) a mechanism
  for inducing inter-molecular quantum entanglement of nuclear spins,
  (iii) a new isotope fractionation mechanism, (iv) a novel
  explanation of the enhanced chemical activity of ``Reactive Oxygen Species'',
  (v) illuminating the importance of ortho-water molecules in modulating the
  quantum dynamics of liquid water, (vi) providing the critical
  quantum-to-biochemical linkage in the nuclear spin model of the
  (putative) quantum brain, among others.
 
\end{abstract}
\pacs{}

\maketitle

\section{Introduction}

\subsection{Motivation}
The far reaching impact of quantum indistinguishability in
few-particle collisions, in molecular spectroscopy\cite{MeadRMP} and
in the low temperature behavior of macroscopic many-body systems
(e.g., superfluidity) is well appreciated and extensively
studied.\cite{LandauQM} However, the role of indistinguishability for
the dynamics of macroscopic systems at high temperature remains
virtually unexplored, typically neglected due to the presumed absence
of necessary quantum coherence. For cohesion of both solids and
molecules, while electrons are treated quantum-mechanically, the much
heavier ions are treated as distinguishable and classical. Moreover,
in chemical reactions of molecules in solution, nuclear spins are
generally believed to play little role, despite their macroscopic
quantum coherence times (especially for spin-$1/2$
nuclei\cite{NMR_Hore,MPAF_AOP}).  However, for small symmetric
molecules the Pauli principle can inextricably entangle the coherent
nuclear spin dynamics with the molecular rotational properties.  The
latter must modulate chemical reaction rates, even if weakly, thereby
coupling nuclear spin dynamics to quantum chemistry.

Molecular hydrogen offers the simplest setting for discussing the
interplay of indistinguishability and chemical reactivity.  While the
two electrons are tightly bound in a symmetric molecular orbital, the
proton nuclear spins are weakly coupled, so that molecular hydrogen
comes in two isomers, para-hydrogen (nuclear spin singlet) and
ortho-hydrogen (nuclear spin triplet).  Treating the motion of the
nuclei quantum-mechanically, the Pauli principle dictates that
molecular para- and ortho-hydrogen rotate with even and odd angular
momentum, respectively.

A natural question, that to the best of our knowledge has not been
asked, is whether the para- and ortho-spin-isomers of molecular
hydrogen exhibit different chemical reaction rates in solvent.  If
yes, as might be expected from the different rotational properties of
the two spin-isomers, what is the magnitude and ``sign'' of the
effect?  Intuitively, one might expect such effects to be small,
especially at temperatures well above the rotational constant.


\subsection{``Quantum Dynamical Selection''}

In this paper we explore this and related questions in a number of
systems, focusing on enzymatic reactions with the substrate consisting
of a small symmetric molecule, characterized by a
``quasi-angular-momentum'', $L_\text{quasi}$ (to be defined in Section
\ref{sec:formulation}). As we elaborate and motivate in the next
section, and in contrast to the aforementioned
``conventional-wisdom'', such bond-breaking chemical reactions can be
very sensitive to nuclear spin states, via Pauli transduction through
the allowed molecular rotations. Our central conjecture is that
symmetric molecules can only have a direct bond-breaking chemical
reaction from a state with an orbitally symmetric wavefunction, i.e.,
with zero `quasi-angular-momentum'' $L_\text{quasi}=0$, (e.g., the
symmetric para-hydrogen). On the other hand, a molecule constrained by
Fermi/Bose indistinguishability to have a nonzero odd orbital angular
momentum is precluded from breaking its bond by a ``Quantum Dynamical
Selection'' (QDS) rule, a ``closed bottleneck'' in Hilbert space.
Physically, this closed bottleneck is due to a destructive
interference between the multiple possible bond-breaking processes --
one for each of the symmetry related molecular orbital
configurations. We emphasize that {\em QDS is not of energetic
  origin}, operational even if the molecule's rotational constant is
much smaller than the temperature.

\subsection{Quantum coherence}
Before exploring the role of quantum indistinguishability on chemical
processes, we briefly comment on the important issue of quantum
decoherence, the common prejudice being that rapid decoherence in a
wet solution will render {\it all} quantum phenomena inoperative.
Indeed, elevated temperatures will generally move a system towards
classical behavior.  For example, when a physical process oscillating
with a characteristic frequency, $\omega$, is immersed in a thermal
environment, quantum effects will typically wash out for T$\ge \hbar
\omega/k_B$.  At body/room temperature it is thus only phenomena
oscillating at very high frequencies ($10^{13} \text{Hz}$, say, such
as molecular vibrational modes) where quantum mechanics can modify the
dynamics.  But this argument implicitly presumes thermal equilibrium.

Nuclei with spin-$1/2$ in molecules or ions tumbling in water are so
weakly coupled to the solvent that macroscopic coherence times of
seconds or minutes are possible and regularly measured in liquid state
NMR.\cite{NMR} But weak coupling is a two-way street; if the solvent
disturbs only weakly the nuclear spins, the nuclear spin dynamics will
only weakly disturb the dynamics of the molecule and the surrounding
solvent. However, small {\it symmetric} molecules, where quantum
indistinguishability can entangle nuclear spin states with molecular
rotations, provide an exception.

As we detail in the next section, the symmetry of the nuclear spin
wavefunction in such symmetric molecules will dictate a characteristic
``quasi-angular-momentum'', $L_\text{quasi}$ -- equal to a small
integer in units of $\hbar$ -- that is symmetry protected even in the
non-rotationally invariant solvent environment.  And, provided the
molecule's thermal angular momentum is much larger than $\hbar$ the
environment can not readily measure $L_\text{quasi}$, so that the
different nuclear-spin symmetry sectors will remain coherent with one
another for exponentially long times.  Remarkably, even though the
solvent is ineffective at measuring $L_\text{quasi}$, we will argue
that enzymes (which catalyze irreversible bond-breaking chemical
reactions) can, in effect, measure $L_\text{quasi}$ -- implementing a
projective measurements onto $L_\text{quasi}=0$.

\subsection{Outline}
The rest of the paper is organized as follows. In Section
\ref{sec:formulation}, focusing on small symmetric molecules with
C$_n$ symmetry, we formulate the general problem and then for
concreteness specialize to the case of $n=2$ and $n=3$.  With this
formulation, in Sec.\ref{sec:QDS} we then state our Quantum Dynamical
Selection conjecture, discussing both physical and mathematical
plausibility arguments for it in Sec.\ref{sec:motivateQDS}. We
conclude in Sec.\ref{sec:experiments} with experimental implications
of the QDS rule, proposing a number of experiments to test the
conjecture.

\section{Theoretical Framework}
\label{sec:formulation}

\subsection{Beyond Born-Oppenheimer approximation}
\label{sec:beyondBO}

In this section we present the basic framework for our subsequent
discussion of the role of symmetry and quantum indistinguishability in
chemical processes involving catalytically-assisted bond-breaking in
symmetric molecules.  In molecular processes the electrons, as fast
degrees of freedom, are appropriately treated as fully
quantum-mechanical indistinguishable fermions.  In contrast, the
constraints of quantum indistinguishability on the nuclear orbital
degrees of freedom when treated within the Born-Oppenheimer
approximation are invariably neglected, especially in solution
chemistry (although not always in molecular
spectroscopy\cite{MeadRMP}). The molecular dynamics and chemical
reactions are thus assumed to be fully controlled by classical motion
of the molecular collective coordinates on the Born-Oppenheimer
adiabatic energy surface.  While this may be adequate for some
systems, we argue that it can be wholly insufficient for chemical
reactions in small symmetric molecules.  As we will discuss, in such
systems, the nuclear and electron spin degrees of freedom can induce
Berry phases that constrain the molecular orbital dynamics on the
adiabatic energy surface, which must then be treated quantum
mechanically since these Berry phases do not enter the classical
equations of motion.  As we shall argue, the presence of Berry phases
can have strong, and previously unappreciated order-one effects in
chemical bond-breaking processes.

\subsection{Planar symmetric molecules}
\label{planarC3molecule}

For simplicity of presentation we focus primarily on molecules which
possess only a single $n$-fold symmetry axis that under a $2\pi/n$
planar rigid-body rotation (implemented by the operator $\hat{C}_n$)
cyclically permutes $n$ indistinguishable fermionic nuclei.  A water
molecule provides a familiar example for $n=2$ and ammonia for $n=3$,
wherein the protons are cyclically permuted.

For such molecules the nuclear spin states can be conveniently chosen
to be eigenstates of $\hat C_n$, acquiring a phase factor (eigenvalue)
$\omega_n^{-\tau}$ with $\omega_n = e^{i 2\pi/n}$ and the
``pseudospin'', $\tau$ taking on values $\tau=0,1,2,...,n-1$.  Due to
Fermi statistics when the molecule is physically rotated by $2\pi/n$
radians around the $C_n$ symmetry axis, the total molecular
wavefunction -- which consists of a product of nuclear rotations,
nuclear spins and electronic molecular states -- must acquire a sign
$(-1)^{n-1}$ due to a cyclic permutation of fermionic nuclei.
Provided the electron wavefunctions transform trivially under the
$C_n$ rotation, this constraint from Fermi statistics of the nuclei,
$e^{i(-2\pi\tau/n + 2\pi L/n - \pi (n-1))} = 1$, implies that the
collective orbital angular momentum of the molecule is constrained by
$\tau$, taking on values $L= L_\text{quasi} + n\mathbb{Z}$ with the
``quasi-angular-momentum'' given by
\begin{eqnarray}
L_\text{quasi}
&=&
\left\{\begin{array}{lr}
\tau,& n\ \  \text{odd}\;,\\
\tau + n/2,& n\ \ \text{even}\;.\\
\end{array}\right.
\label{Lquasi}
\end{eqnarray}
%


\begin{figure}[tbp]
\includegraphics[height=3.5in,width=3.5in,keepaspectratio]{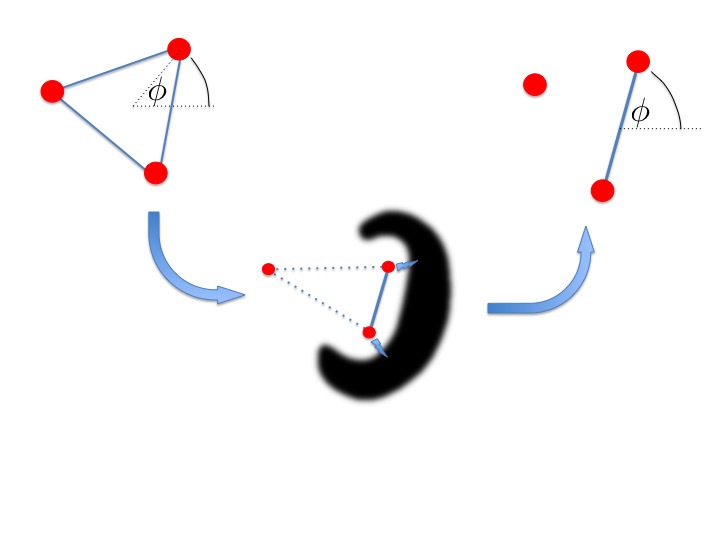}
\caption{Schematic of a bond-breaking chemical reaction: (a) Initial
  state of a $C_3$ symmetric molecule with two molecular electrons
  (blue) bonding the three nuclei (red) together. (b) An intermediate
  state where the reaction is catalyzed by an enzyme that ``grabs''
  two of the nuclei, weakening their bonds to the third by depletion
  of electronic charge. (c) Final product state composed of a
  molecular dimer and an isolated atom.}
\label{fig:molecule3}
\end{figure}

\subsubsection{Molecular trimer of identical fermionic nuclei}

For illustrative clarity we first formulate the problem for the case
of $n=3$, specializing to three identical fermionic nuclei with
nuclear spin $1/2$ and electrical charge $+1$.  We focus on a singly
ionized trimer molecule T$\equiv$A$_3^+$ undergoing a chemical
reaction,
\begin{equation}
\text{A}_3^+ \rightarrow \text{A}_2 + \text{A}^+  ,
\end{equation}
into a singly ionized atom, A$^+$, and a neutral dimer molecule,
D$\equiv $A$_2$. The process is schematically displayed in
Fig.\ref{fig:molecule3}, where the molecular trimer is composed of the
three fermionic nuclei with creation operators,
$\hat{F}^\dagger_\alpha$, in spin state $\alpha =
\uparrow,\downarrow$, that form a C$_3$ symmetric molecular
configuration characterized by a collective coordinate $\phi$.  The
nuclei part of such a molecular state is created by the three-nuclei
operator,
\begin{equation}
\hat{T}^\dagger_\tau (\phi) = \sum_{\alpha \beta\gamma}
\chi^\tau_{\alpha \beta \gamma} \hat{F}_\alpha^\dagger(\phi)
\hat{F}_\beta^\dagger(\phi+ 2\pi/3) \hat{F}_\gamma^\dagger(\phi+
4\pi/3).
\label{Toperator}   
\end{equation}

The three-nuclei spin wavefunction $\chi^\tau_{\alpha \beta \gamma}$
is chosen as a $\tau$ representation of cyclic permutations, an
eigenstate of $\hat{\text{C}}_3$,
\begin{equation}
  \hat{\text{C}}_3\chi^\tau_{\alpha \beta \gamma} =\chi^\tau_{\gamma \alpha
    \beta} = \omega_3^\tau \chi^\tau_{\alpha \beta \gamma},  
\end{equation}
with $ \omega_3 = e^{i 2 \pi /3}$, required by
$(\hat{\text{C}}_3)^3=1$. Thus, for this trimer molecule, $\tau$ takes
on one of three values, $\tau =0,1,2$ (or equivalently, $\tau =0,\pm
1$). By construction $\hat{T}_\tau (\phi)$ then also forms an
irreducible representation of $\hat{\text{C}}_3$ satisfying,
\begin{equation}
\hat{T}_\tau (\phi + 2\pi/3)  = \omega_3^\tau \hat{T}_\tau (\phi).
\end{equation}
We will at times refer to $\tau$ as a ``pseudospin", that encodes both
the nuclear spin and the orbital qubits, entangled through the Pauli
principle of identical nuclei.

Because a discrete $2\pi/3$ rotation executes a fermionic cyclic
interchange, the $\tau$ representation of the nuclear spin
wavefunction imprints a nontrivial Berry phase $\omega^\tau$ onto the
orbital degree of freedom, $\phi$, which we discuss below.  For
simplicity we have suppressed the position coordinate describing the
center of mass of the trimer molecule as well as the orientation of
the normal to this planar trimer molecule.

In addition to the nuclei, a correct description of a molecule must
also consist of bonding electrons that, within the Born-Oppenheimer
approximation, occupy the molecular orbitals.  For concreteness we
consider a (singly ionized) molecule with only two electrons that form
a spin-singlet in the ground-state molecular orbital $\psi_{T} ({\bf
  r}; \phi)$, where the subscript denotes the trimer nuclear
configuration. This wavefunction transforms symmetrically under
$\hat{\text{C}}_3$, satisfying $\psi_{T} ({\bf r}; \phi + 2\pi/3) =
\psi_{T} ({\bf r}; \phi )$. We denote the electron creation operator
in this orbital as,
\begin{equation}
\hat{c}^\dagger_{ \sigma}(\phi) = \int_{ {\bf r}}  \psi_{T}  ({\bf r}; \phi) \hat{c}^\dagger_\sigma({\bf r}).
\end{equation}

The trimer molecular state we thus consider can be written as,
\begin{equation}
 | T \rangle = \sum_{\tau}  \int_\phi \Psi_\tau(\phi) \hat{c}^\dagger_{\uparrow}(\phi) \hat{c}^\dagger_{\downarrow}(\phi) \hat{T}^\dagger_\tau (\phi)  | \text{vac} \rangle \otimes | {\cal E} \rangle_\phi,
\label{Tstate}
\end{equation}
characterized by an orbital wavefunction in the $\tau$ representation,
\begin{equation}
\Psi_\tau(\phi + 2\pi/3) = \omega_3^\tau \Psi_\tau(\phi).
\label{Psi_tau}
\end{equation}
Were the general orbital wavefunction $\Psi_\tau(\phi)$ expanded in
angular momentum eigenstates, $e^{iL \phi}$, with $L \in \mathbb{Z}$,
this constraint implies that $L = L_\text{quasi} + 3 \mathbb{Z}$ with
a ``quasi-angular-momentum'' $L_\text{quasi}=\tau$ , consistent with
Eq.\rf{Lquasi} for $n=3$.

In Eq.\rf{Tstate} the ket $|{\cal E} \rangle_\phi$ denotes the initial
quantum state of the environment -- i.e., the solvent and enzyme --
that is entangled with the molecular rotations through the angle
$\phi$, as generically the environment can ``measure'' the molecular
orientation.  Note that we have implicitly assumed that the initial
state of the environment does not depend on $\tau$, so that $|{\cal E}
\rangle_{\phi + 2\pi/3} = |{\cal E} \rangle_{\phi}$.  For a molecule
with thermal angular momentum, $L_T$ (defined through $\hbar^2 L_T^2 /
{\cal I} = k_B T$, with moment of inertia ${\cal I}$) that is much
greater than one, $L_T \gg 1$, the solvent is ineffective in
``measuring'' the molecular quasi-angular-momentum, which is a small
fraction of $L_T$.  Indeed, the quasi-angular-momentum decoherence
time should be exponentially long for large thermal angular momentum,
varying as $t_{\text{coh}}^\tau \sim t_0 \exp(c L_T^2)$ with an order
one constant $c$ and $t_0$ a microscopic time of order a picosecond.

\subsubsection{Molecular dimer and atom products state}

As illustrated in Fig.\ref{fig:molecule3}, the bond-breaking reaction
proceeds through an intermediate enzymatic stage, that is challenging
to describe microscopically. Through its interaction with the
electronic orbital degrees of freedom the enzyme temporarily binds and
``holds'' two of the nuclei, separating them from the third nucleus.
This causes molecular rotations to cease and also weakens the
molecular bonds.  We assume that the final product state consists of a
neutral molecular dimer, D$\equiv \text{A}_2$, held together by the
two electrons and a singly ionized atom A$^+$.  The orientation of the
dimer is characterized by a single angle coordinate, which we again
denote as $\phi$ (see Figure \ref{fig:molecule3}).

This final product state $|P\rangle$ can then be expressed as,
\begin{equation}
|P\rangle = \sum_\tau a_\tau \int_\phi  \sum_{\mu, \alpha \beta\gamma} \Psi_{\mu}({\phi}) 
\chi^\tau_{\alpha \beta \gamma}|A_\alpha\rangle
|D_{\beta\gamma},\phi \rangle \otimes | {\cal E}^* \rangle_\phi,
\label{productState}
\end{equation}
where $| {\cal E}^* \rangle_\phi$ describes the state of the
environment (solvent plus enzyme) after the chemical reaction,
$|A_\alpha \rangle = \hat{F}_{\rv_A,\alpha}^\dagger |\text{vac}
\rangle$ denotes the state of the atom (located at position $\rv_A$),
and the state of the dimer molecule (located at position $\rv_D$) is
given by,
 \begin{equation}
  |D_{\beta\gamma}, \phi \rangle = 
\hat{F}_{\rv_D,\beta}^\dagger(\phi)
\hat{F}_{\rv_D,\gamma}^\dagger(\phi+\pi)
\hat{c}^\dagger_{ \uparrow}(\phi) \hat{c}^\dagger_{ \downarrow}(\phi) 
  |\text{vac} \rangle .
\end{equation}
The electron creation operators on the dimer molecule are given by,
 \begin{equation}
\hat{c}^\dagger_{ \sigma}(\phi) = \int_{ {\bf r}}  \psi_{D}  ({\bf r}; \phi) \hat{c}^\dagger_\sigma({\bf r}) ,
\end{equation}
with the ground state molecular orbital for the dimer molecule (when
oriented at angle $\phi$) $\psi_{D} ({\bf r}; \phi)$, assumed to
transform symmetrically under the $\text{C}_2$ symmetry of the dimer
molecule, $\psi_{D} ({\bf r}; \phi + \pi) = \psi_{D} ({\bf r}; \phi)$.
The dimer orbital wavefunction $\Psi_\mu({\phi})$ transforms as
$\Psi_\mu({\phi} + \pi) = e^{i \mu \pi} \Psi_\mu({\phi})$ with $\mu =
0,1$.

Because we do not expect the nuclear spins state in each $\tau$-sector
to change through the chemical reaction, the atomic and dimer states
remain entangled through the nuclear spin wavefunction,
$\chi^\tau_{\alpha \beta \gamma}$.  We have introduced an overall
amplitude, $a_\tau$ which we shall discuss further below.

Since an enzymatic chemical reaction will typically be strongly
exothermic (releasing, for example, a fraction of eV in energy) the
quantum state of the environment after the reaction, $|{\cal E}^*
\rangle_\phi$ will be very different than before the chemical
reaction, $|{\cal E} \rangle_\phi$ -- that is $_\phi\langle {\cal E}^*
|{\cal E} \rangle_\phi = 0$.

\subsubsection{Generalization to arbitrary n-mer}

Here we briefly generalize from $n=3$ to a planar molecule with
$n$-fold symmetry consisting of $n$ fermionic spin-$1/2$ nuclei.  The
nuclear spin wavefunction $\chi^{\tau}_{\alpha_1 \alpha_2
  ... \alpha_n}$ can be chosen as an eigenstate of the cyclic
permutation symmetry, $\hat{\text{C}}_n$,
\begin{equation}
\hat{\text{C}}_n \chi^{\tau}_{\alpha_1 \alpha_2 ... \alpha_n} =  \chi^{\tau}_{\alpha_n \alpha_1 \alpha_2 ... \alpha_{n-1}}  
 = \omega_n^{\tau} \chi^{\tau}_{\alpha_1 \alpha_2 ... \alpha_n}  ,
\end{equation}
with $ \omega_n = e^{i 2 \pi /n}$ required by $(\hat C_n)^n=1$.  The
pseudospin now takes on one of $n$ values, $\tau = 0,1,2,...n-1$.  Due
to Fermi statistics of the nuclei the total molecular wavefunction
must acquire a factor of $(-1)^{n-1}$ under a molecular rotation by $2
\pi/n$.  Assuming that the bonding electrons transform trivially under
$\text{C}_n$, the molecular orbital wavefunction (as in
Eq.\rf{Psi_tau}) must satisfy, $\Psi_{\tau}(\phi + 2\pi/n) =
(-1)^{(n-1)} \omega_n^{\tau} \Psi_{\tau}(\phi)$.  Equivalently, the
allowed orbital angular momenta are given by $L=L_\text{quasi} + n
\mathbb{Z}$ with the quasi-angular-momentum $L_\text{quasi}$ given in
Eq.(\ref{Lquasi}).

\section{Quantum Dynamical Selection Rule}
\label{sec:QDS}

We can now state our conjecture, which we refer to as a ``Quantum
Dynamical Selection'' (QDS) rule:

{\it A bond-breaking enzymatic chemical reaction on a symmetric planar
  molecule implements a projective measurement onto zero quasi-angular
  momentum}, $L_\text{quasi} = 0$.

More generally, including for molecules with three-dimensional
rotational symmetries such as $\text{H}_2$ and $\text{CH}_4$, our
Quantum Dynamical Selection rule implies that:

{\it Enzymatic chemical
  reactions that (directly) break the bonds of a symmetric molecule
  are strictly forbidden from orbitally non-symmetric molecular
  states.}  

Here, ``direct'' implies that the transition proceeds without the
molecule first undergoing a nuclear spin flip.  For example, the
ortho-state of molecular hydrogen (which has odd angular momentum)
cannot undergo a direct bond-breaking transition without passing
through the para-state.  For the $n=3$ planar molecule described in
Section \ref{planarC3molecule} our QDS rule implies that the amplitude
in Eq.(\ref{productState}) vanishes unless $\tau=0$, that is $a_\tau= \delta_{\tau 0}$.  In the following Section we present an argument that offers
support for the Quantum Dynamical Selection rule.

\section{Arguments for QDS conjecture}
\label{sec:motivateQDS}

For conceptual reasons it is helpful to divide the enzyme mediated
chemical reaction into three stages, each a distinct quantum state of
the molecule and enzyme: (a) a state $\psi_a$, in which the molecule
is free to rotate and dynamical processes that exchange the atoms are
allowed, (b) a state $\psi_b$, in which the molecule's rotations are
stopped by the enzyme and dynamical processes that exchange the atoms
are forbidden, and (c) a state $\psi_c$, in which the chemical bond is
broken and the molecule is fragmented into its constituents.  For each
of these three quantum states a careful discussion of the properties
of the accessible Hilbert space will be necessary, and is taken up in
the first three subsections below.

The full enzymatic chemical reaction corresponds to a dynamical
process that takes the system from state $\psi_a$ to $\psi_b$ and then
to $\psi_c$.  In subsection D, below, we explore the tunneling
amplitude between these quantum states.  Since these transitions are
either microscopically or macroscopically irreversible the full
enzymatic reaction should be viewed as implementing ``projective
measurements'' on the molecule.  As we will demonstrate, the tunneling
amplitude vanishes unless $L_\text{quasi}=0$, which offers an argument
for the validity of the conjectured Quantum Dynamical Selection rule.

\subsection{A rotating symmetric molecule}

\begin{figure}[tbp]
\includegraphics[height=4in,width=3.5in,keepaspectratio]{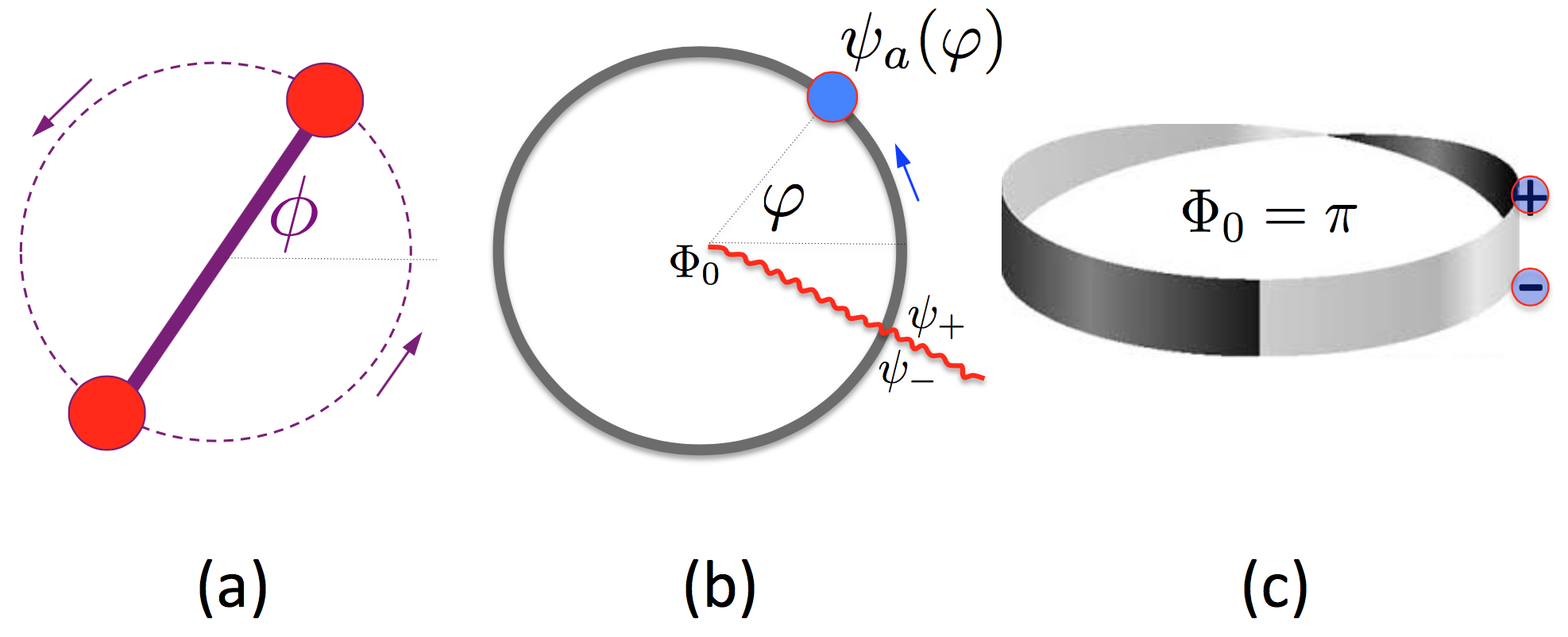}
\caption{Three representations of a rotating diatomic molecule
  composed of identical nuclei: (a) An explicit physical
  representation.  (b) An effective model for the dynamics of the
  angular coordinate, $\varphi = 2\phi\in [0, 2\pi )$, for a molecule
  with $C_2$ symmetry, represented as a quantum bead on a ring with
  Berry flux, $\Phi_0$ determined by the nuclear spin wavefunction.
  The bead wavefunction, $\psi_a(\varphi)$, is discontinuous across an
  n-th root branch cut (red squiggly line), $\psi_+ = e^{i \Phi_0}
  \psi_-$. (c) Placing the bead on an n-fold cover of the ring with
  $\varphi \in [0,2\pi n)$ -- a Mobius strip for the $n=2$
  double-cover -- resolves any ambiguity in the placement of the
  branch cut.}
\label{RotatingMoleculeFig}
\end{figure}

We begin with a precise definition of the initial quantum mechanical
state of the rotating symmetric molecule, offering three
representations of the accessible Hilbert space. For the case of a
diatomic ($n=2$) molecule, these are illustrated in
Fig.\ref{RotatingMoleculeFig}.


\subsubsection{Mapping to quantum bead on a ring}

To proceed requires a careful discussion of the Hilbert space for the
``angle'' kets, that are eigenkets of the angle operator, $\hat{\phi}
| \phi \rangle = \phi | \phi \rangle$. They are defined (for $n=3$) as
$| \phi \rangle = \hat{T}_\tau (\phi) | \text{vac} \rangle$, where
$\hat{T}_\tau (\phi)$ was introduced in Eq.\rf{Toperator}, with a
natural generalization for all $n$.  For the $n$-fold planar molecule
we have,
\begin{equation}
| \phi + 2\pi/n \rangle = e^{i \Phi_0}  | \phi \rangle,
\end{equation}
with 
\begin{equation}
\frac{\Phi_0}{2\pi} = \frac{L_\text{quasi}}{n}   ,
\end{equation}
so that the angle kets are redundant on the full interval $[0,2\pi)$.
As such, it is convenient to restrict the angle $\phi \in [0,2\pi
/n)$.  Then, we can define a new angle operator $\hat{\varphi} \in [0,
2\pi)$ and its canonically conjugate angular-momentum operator
$\hat{\ell} \in \mathbb{Z}$ with $[\hat{\varphi}, \hat{\ell} ] = i$
via,
\begin{equation}
\hat{\phi} = \hat{\varphi}/n    ;  \hskip0.4cm  \hat{L} = n \hat{\ell} + L_\text{quasi}   .
\end{equation}

Consider the simplest Hamiltonian for a rotating molecule,
\begin{equation}
\hat{H}_n = \frac{  \hat{L}^2}{2 {\cal I}_n} + V_n(\hat{\phi}) ,
\end{equation}
with $V_n(\phi + 2 \pi/n) = V_n(\phi)$ an environmental potential
(e.g., the enzyme, solution, etc.), with its periodicity encoding the
C$_n$ symmetry of the molecule. Re-expressing this Hamiltonian in
terms of the ``reduced'' variables, $\hat{H}_n \rightarrow {\cal
  \hat{H}}$ gives,
\begin{equation}
\hat{{\cal H}} = \frac{ ( \hat{\ell} + \Phi_0/2\pi )^2}{2 {\cal I}} + U( \hat{\varphi}) ,
\end{equation}
with a $2\pi$-periodic potential, $U(\varphi + 2\pi) =
U(\varphi)\equiv V_n(\varphi/n)$ and a rescaled moment of inertia,
${\cal I} = n^2 {\cal I}_n$.  This Hamiltonian can be viewed as
describing a ``fictitious'' quantum bead on a ring, with a
``fictitious'' magnetic flux, $\Phi_0$, piercing the ring, as shown in
Fig.\ref{RotatingMoleculeFig}.

To expose the physics of this flux it is useful to consider the
Lagrangian of the quantum bead on the ring:
\begin{equation}
{\cal L} = \frac{1}{2} {\cal I} (\partial_t \varphi)^2 - U (\varphi) + {\cal L}_B ,
\end{equation}
with Berry phase term, ${\cal L}_B = (\Phi_0/2\pi) ( \partial_t
\varphi)$.  In a real (or imaginary) time path integral this Berry
phase term contributes an overall multiplicative phase factor,
\begin{equation}
e^{iS_B} = e^{i \Phi_0 W},
\end{equation}
with $W \in \mathbb{Z} $ a winding number defined via $2\pi W =
\varphi(t_f) - \varphi(t_i)$, where $t_i,t_f$ are the initial and
final times, respectively.

It must be emphasized that the wavefunction for the bead on the ring,
which we denote as $\psi_a(\varphi)$, is not single valued for
$L_\text{quasi}\neq 0$, since $\psi_a({\varphi + 2\pi) = e^{i \Phi_0}
  \psi_a(\varphi})$, so that a branch-cut is required.  For the planar
molecule with C$_n$ symmetry this will be an $n^{th}$-root of unity
branch cut, while for the ortho-dimer molecule it is a square-root
cut.  In either case, the wavefunction across the branch cut is
discontinuous, with the value on either side of the branch cut,
denoted $\psi_+, \psi_-$, being related by a phase factor, $\psi_+ =
e^{i \Phi_0} \psi_-$ (see Fig.\ref{RotatingMoleculeFig}(b)). For an
isolated molecule this branch cut can be placed anywhere (gauge
invariance) but during the enzymatic bond breaking process a natural
gauge invariant formulation is not readily apparent.

\subsubsection{Mobius strip and Umbilic Torus}

In order to resolve any ambiguity in the placement of the branch cut
with bond breaking present, it is helpful to view the bead as living
on an {\it n-fold cover of the ring}.  Mathematically, we simply
extend the range of $\varphi$ to lie in the interval, $\varphi \in
[0,2\pi n)$, so that the wavefunction is periodic in this enlarged
domain, $\psi_a(\varphi + 2\pi n) = \psi_a(\varphi)$.  For $n=2$ with
$L_\text{quasi}=1$ this corresponds to a quantum bead living on the
(single) ``edge'' of a Mobius strip, as depicted in
Fig.\ref{RotatingMoleculeFig}(c).  At a given angle on the Mobius
strip the wavefunction on the opposite edges of the strip must have a
sign change, since in this case $\psi_a(\varphi + 2\pi) = -
\psi_a(\varphi)$.  For $n=3$ with $L_\text{quasi} = \pm 1$ the quantum
bead lives on a (3-sided) umbilic torus, which must be circumnavigated
three times before returning to the same ``edge''.

\subsection{A ``not-rotating''  molecule}

A first step in the bond-breaking process requires stopping of the
molecule's rotation as it binds to the enzyme.  What does it mean for
a small symmetric molecule to be ``not-rotating''?  Perhaps having
zero angular velocity?  For 1d translational motion the linear (group)
velocity of a quantum particle is $v_g = \partial E_p / \partial p$,
suggesting that angular (group) velocity should be likewise defined,
$\Omega = \partial E_L / \partial L$.  But angular momentum is
quantized in units of $\hbar$, so this definition is problematic.

One plausible definition of a molecule to be ``not-rotating'' is for
its orbital motion to be described by a real wavefunciton.  For odd
$n$ this is equivalent to requiring zero quasi-angular-momentum,
$L_\text{quasi}=0$, while for $n$-even
it is not since for $L_\text{quasi}=n/2 \ne 0$ a
real wavefunction can still be constructed.
But in either case, once
we allow for quantum entanglement between the molecule and the
solvent/enzyme the notion of the ``molecule's wavefunction'' becomes
problematic.

\begin{figure}[tbp]
\includegraphics[height=3.3in,width=3.3in,keepaspectratio]{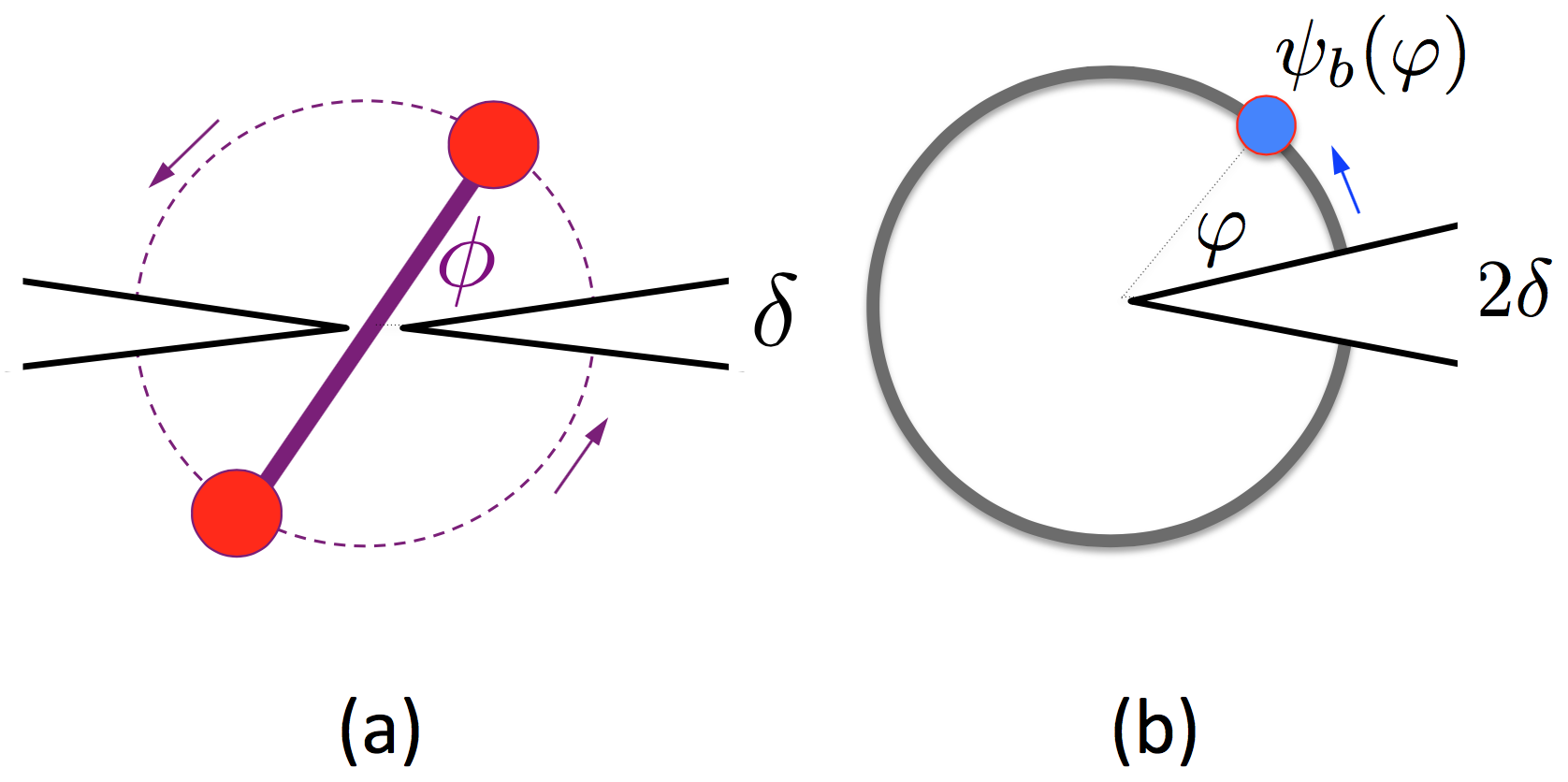}
\caption{Two representations of a {\em non-rotating} diatomic molecule
  composed of identical nuclei: (a) An explicit physical
  representation with impenetrable potential wedges inserted around
  $\phi = 0$ and $\pi$, that restrict the molecular rotation angle and
  constrain one nuclei to the upper-half plane, and the other to the
  lower half-plane.  These wedges strictly forbid all rotational
  motions that dynamically exchange the two nuclei. The {\em
    identical} nuclei thus become effectively {\em distinguishable}.
  (b) An effective quantum bead model, where the ``ring'', now
  restricted to the angular range, $\varphi \in [\delta,
  2\pi-\delta)$, has been ``cut-open'' - and the bead is constrained
  to move on a line-segment.}
\label{nonRotatingMoleculeFig}
\end{figure}

We believe the best way to impose a ``not-rotating'' restriction of
the molecule, is to impose a constraint that disallows a dynamical
rotation which implements an exchange of the constituent atoms.  This
can be achieved by inserting impenetrable potential wedges at angles
centered around $\phi = 0$ and $\phi = \pi$, as illustrated in
Fig. \ref{nonRotatingMoleculeFig}(a) for the diatomic molecule.  These
wedges restrict the molecular rotation angle so that one of the nuclei
is restricted to the upper-half plane, $\delta/2 < \phi \le
\pi-\delta/2$, while the other resides in the lower half-plane,
$\pi+\delta/2 < \phi < 2\pi-\delta/2$.  Rotations that exchange the
nuclei are thereby strictly forbidden.  And the two nuclei, while
still {\em identical}, have effectively become ``{\em
  distinguishable}'' - even if their spins are aligned.

Upon mapping to the effective coordinate, $\phi \rightarrow
\varphi/n$, the wavefunction of the bead on the ring in this
``not-rotating'' configuration, which we denote as $\psi_b(\varphi)$,
is restricted to the {\em line segment} $\varphi \in (\delta,
2\pi-\delta)$.  As illustrated in Fig.\ref{nonRotatingMoleculeFig}(b),
the ``ring'' has in effect been ``cut-open''.  As such, there is no
meaning to be ascribed to the effective flux through the ring, present
in Fig.\ref{RotatingMoleculeFig} for the rotating molecule.  In
contrast to the n-fold cover of the rotating wavefunction,
$\psi_a(\varphi + 2\pi n) = \psi_a(\varphi)$, the non-rotating
wavefunction, $\psi_b(\varphi)$ is defined on a single cover of the
cut-open ring with $\varphi \in (\delta, 2\pi-\delta)$.
 
\subsection{A bond-broken molecule}

\begin{figure}[tbp]
\includegraphics[height=3.3in,width=3.3in,keepaspectratio]{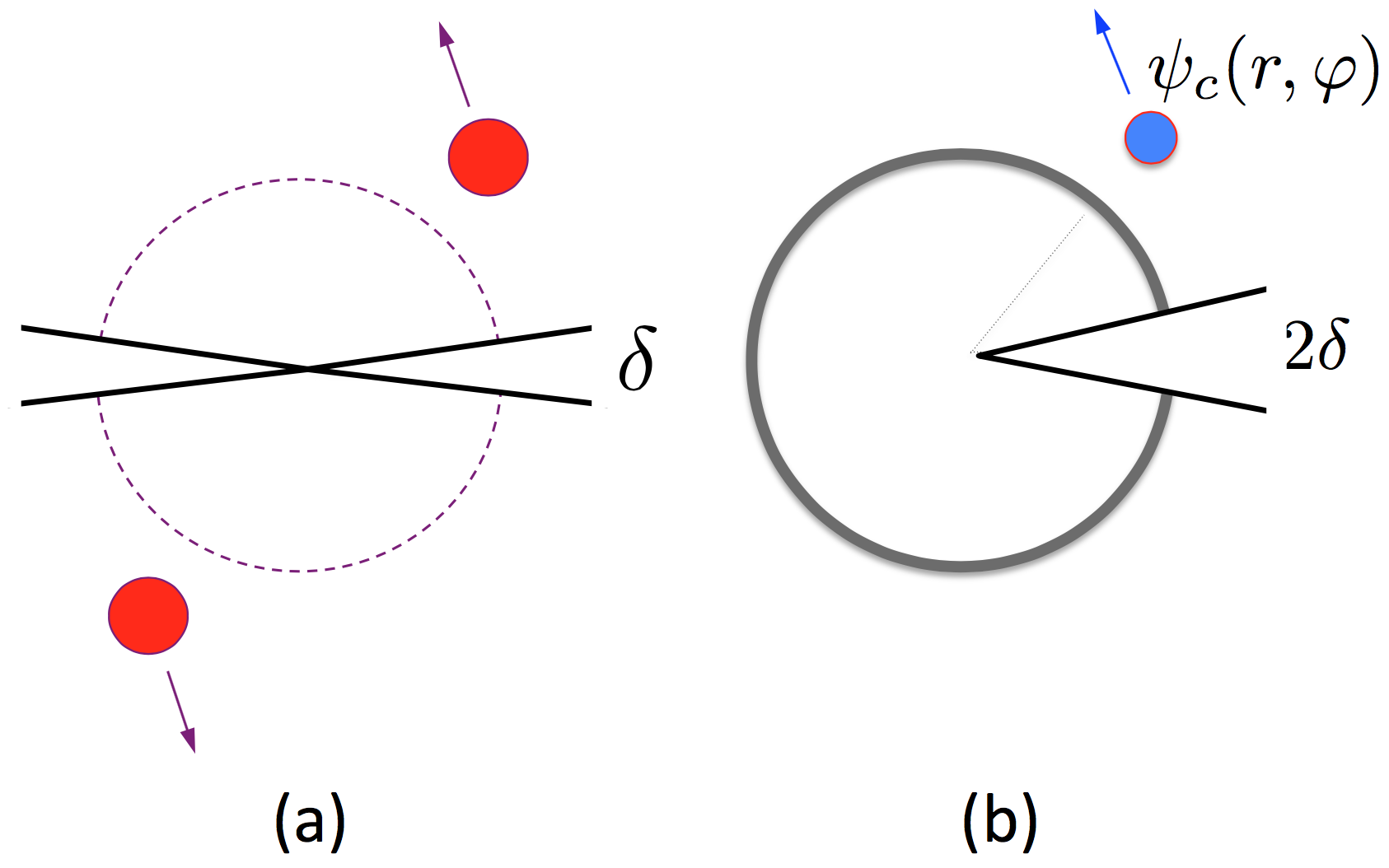}
\caption{Two representations of a bond-broken diatomic molecule
  composed of identical but {\em distinguishable} nuclei: (a) An
  explicit physical representation with impenetrable potential wedges
  inserted at $\phi = 0$ and $\pi$, allowing for bond-breaking process
  to the distinguishable-nuclei state. (b) An effective quantum bead
  on a ring model, restricted to the angular coordinate range,
  $\varphi \in [\delta, 2\pi-\delta)$, with bond-breaking modeled by
  tunneling the bead off the ring.}
\label{BondBreakingFig}
\end{figure}

Once the molecule is ``not-rotating'' and the identical fermions are
effectively ``distinguishable'', a breaking of the chemical bond is
greatly simplified.  As illustrated in Fig.\ref{BondBreakingFig}(a),
with the impenetrable barriers present the molecular bond can break
and the constituent atoms are free to move off into the upper and
lower half-planes, respectively.  Once the atoms are physically well
separated, their ``distinguishability" no longer rests on the presence
of the impenetrable barriers.

In the quantum-bead representation, illustrated in
Fig.\ref{BondBreakingFig}(b), the bond-breaking process corresponds to
the bead tunneling off the line-segment.  As such, the location of the
bead must now be specified by both the angle $\varphi$ and a radial coordinate, $r$,
the bead wavefunction taking the form $\psi_c(r,\varphi)$.

Since this bond-breaking process will typically be macroscopically
irreversible we assume that once the bead tunnels off the line
segment it will not return.  The tunneling rate for this
process can be then expressed in terms of a Fermi's Golden rule, $\Gamma_{b
  \rightarrow c} \sim |{\cal A}_{b \rightarrow c} |^2$, with tunneling
amplitude,
\begin{equation}
  {\cal A}_{b \rightarrow c}  = \int_\delta^{2 \pi - \delta} d \varphi  \hskip0.1cm H_{bc}(\varphi)  \psi^*_c(R,\varphi) \psi_b(\varphi)  ,
\end{equation}
with $H_{bc}$ a tunneling Hamiltonian and $R$ the radius of the
molecule.

\subsection{Indistinguishable-to-Distinguishable Projective Measurement}

We now turn to the more subtle process where the ``rotating'' molecule
transitions into the ``not-rotating'' state, which occurs when the
enzyme ``catches" the molecule.  When the molecule is rotating, in
state $\psi_a$, dynamical processes that interchange the identical
fermions are allowed and will generally be present.  As such, these
identical fermions are truly ``indistinguishable'' in the rotating
state.  But when the enzyme catches and holds the molecule in place,
in the state $\psi_b$, a projective measurement of the atomic
positions has been implemented (the enzyme is the ``observer'') and the
identical fermions are now ``distinguished''.  

The transition rate for this process, $\Gamma_{a \rightarrow b}$, can be expressed as a product of a microscopic attempt frequency, $\omega_{ab}$ and a Born measurement probability, $P_{ab}$, that is $\Gamma_{a \rightarrow b} = \omega_{ab} P_{ab}$.  The Born probability can in turn be expressed as the squared overlap of the projection of the ``distinguishable" state, $\psi_b$, onto the ``indistinguishable" state, $\psi_a$, that is
$P_{ab} = | \langle \psi_b | \psi_a \rangle |^2$.


Since the rotating molecule wavefunction, $\psi_a(\varphi)$, is
defined on the n-fold cover of the ring (Mobius strip for $n$=2) with
$\varphi \in (0, 2\pi n)$, while the not-rotating wavefunction,
$\psi_b(\varphi)$ lives on a single cover of the line segment,
$\varphi \in (\delta, 2\pi - \delta)$, the two states are seemingly
defined in a different Hilbert space.  This ambiguity
can be resolved by noting that tunneling from the Mobius strip onto
the open line segment can occur from any of the $n$ ``edges" of the
Mobius strip, and that these $n$ processes must be summed over,
\begin{equation} 
\langle \psi_b | \psi_a \rangle  = \sum_{m=0}^{n-1}
  \int_\delta^{2 \pi - \delta} d \varphi \hskip0.1cm 
  \psi^*_b(\varphi) \psi_a(\varphi + 2\pi m) .
\end{equation}
Upon using the $2 \pi$ periodicity condition for the Mobius strip
wavefunction, $\psi_a(\varphi + 2\pi m) = e^{im \Phi_0}
\psi_a(\varphi)$ with $\Phi_0 = 2\pi L_\text{quasi} /n$, this can be
re-expressed as,
\begin{equation} 
\langle \psi_b | \psi_a \rangle = {\cal C}_n
  \int_\delta^{2 \pi - \delta} d \varphi \hskip0.1cm 
  \psi^*_b(\varphi) \psi_a(\varphi) ,
\end{equation}
with the amplitude,
\begin{equation}
{\cal C}_n = \sum_{m=0}^{n-1} e^{i m \Phi_0}
 = \sum_{m=0}^{n-1}  e^{i 2\pi m L_\text{quasi}/n} =  n \delta_{L_\text{quasi},0} .
\end{equation}
The Born probability $P_{ab} = | \langle \psi_b | \psi_a \rangle |^2$ thus vanishes unless $L_\text{quasi}=0$, as does the rate, $\Gamma_{a \rightarrow b}$, that the enzyme grabs and holds the molecule in a non-rotating
``distinguishable" configuration. 

\begin{figure}[tbp]
\includegraphics[height=3.4in,width=3.4in,keepaspectratio]{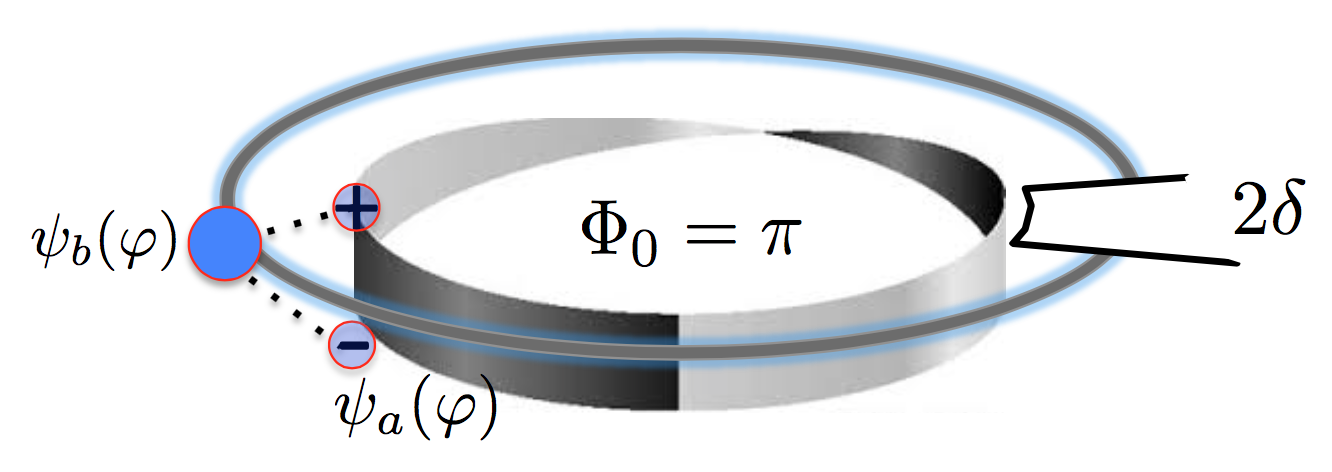}
\caption{The enzymatic projective measurement that induces a
  transition from a ``rotating'' to a ``not-rotating'' molecular state
  can be described as a projective overlap between the rotating state with the
  quantum bead on the Mobius strip, $\psi_a$, and a not-rotating state
  where the bead is on the ``cut-open" outer ring, $\psi_b$.  This
  projective measurement implements a transition between an initial
  state in which the identical atoms are ``indistinguishable: to a
  final ``distinguishable'' state - the enzyme acting as an
  ``observer''.}
\label{IndistingDistingTunnelingFig}
\end{figure}

Physically, for $L_\text{quasi} \ne 0$, there is a destructive
interference between the $n$-parallel paths, each tunneling from an ``edge" of the Mobius strip/torus, onto the line segment.  For the diatomic molecule this is illustrated in Fig.\ref{IndistingDistingTunnelingFig}, where the two tunneling paths with opposite signs will destructively interfere.
The rate for this process which stops the molecule's
rotation, $\Gamma_{a \rightarrow b}$, will vanish unless
$L_\text{quasi}=0$ - the molecule simply can not get caught by the enzyme for non-zero $L_\text{quasi}$,

Since the enzymatic reaction requires both stopping the molecules
rotation, with rate $\Gamma_{a \rightarrow b}$, and subsequently 
breaking the chemical bond, with rate $\Gamma_{b \rightarrow c} \ne
0$, the full chemical reaction rate is,
\begin{equation}
\Gamma_{a \rightarrow c} = \frac{\Gamma_{a \rightarrow b} \Gamma_{b \rightarrow c} } {\Gamma_{a \rightarrow b} + \Gamma_{b \rightarrow c} } \sim \delta_{L_\text{quasi},0} ,
\end{equation}
and vanishes unless $L_\text{quasi} =0$.  Mathematically, the chemical
bond breaking process is ``blocked" by the presence of the Berry's
phase, operative whenever the orbital molecular state is
non-symmetric.  

More physically, this blocking is due to the
destructive interference between the $n$-possible bond-breaking
processes - one for each of the symmetry related molecular orbital
configurations. 
These considerations thus provide support for our
conjectured Quantum Dynamical Selection rule that states the
impossibility of (directly) breaking a chemical bond of a symmetric
molecule rotating non-symmetrically.

\section{Conceptual and experimental implications}
\label{sec:experiments} 

There are numerous experimental implications of the QDS rule.  Since
this rule should indeed be viewed as a conjecture, it will have to be
validated or falsified by comparison of theoretical predictions with
experiments.  Below, we discuss some implications of QDS.

\subsection{Differential reactivity of para/ortho-hydrogen}

Hydrogen provides the most familiar example of molecular spin-isomers,
para-hydrogen with singlet entangled proton spins and rotating with
even angular momentum, and ortho-hydrogen with triplet spin
entanglement and odd rotational angular momentum.  Since the allowed
rotational angular momentum of such homonuclear dimer molecules with
S=$1/2$ fermionic ions (protons) are given by $L= L_\text{quasi} +
2\mathbb{Z}$, the quasi-angular-momentum, while zero for
para-hydrogen, is equal to one for ortho-hydrogen.  The presence of
the Berry phase in the rotation of ortho-hydrogen will suppress the
bond-breaking chemical reactivity.

Many microbes in biology\cite{bioMetaboliteH2} use H$_2$ as a
metabolite, and the enzyme hydrogenase catalyzes the bond-breaking
chemical reaction, H$_2 \rightarrow 2 \text{H}^+ + 2 e^-$.  Based on
the Quantum Dynamical Selection rule we would expect a differential
reactivity between para- and ortho-hydrogen, with the reaction rate
suppressed for ortho-hydrogen.  Indeed, if this reaction were to
proceed ``directly" - without an ionized intermediary or a flipping of
nuclear spin - QDS would predict a complete blocking of ortho-hydrogen
reacting.  At body temperature in a thermal distribution the
ortho:para ratio approaches 3:1 (set by triplet degeneracy), while it
is possible to prepare purified para-hydrogen where this ratio is
strongly inverted (say 1:10).  One might then hope to observe
different enzymatic activity for hydrogenase catalysis in these two
situations, with purified para-hydrogen being significantly more
reactive.

Possible differential combustion of para- and ortho-hydrogen with, say
oxygen, might also be interesting to explore, even though this
reaction is not enzymatic.

\subsection{Intermolecular entanglement of nuclear spins}
\label{sec:spinEntanglement} 

The ability to prepare purified para-hydrogen molecules in solvent and
drive a bond-breaking chemical reaction enables the preparation of two
protons with nuclear spins entangled in a singlet.  If/when these two
protons bind onto a large molecule with different chemical
environments, it is sometimes possible to perform a $\pi$ rotation on
one of the two nuclei to create alignment of the two spins, termed
hyperpolarization.  These hyperpolarized proton spins can then be used
to transfer spin polarization to the nuclei of atoms on the molecules
to which they are bonded.

There is, of course, a long precedent for liquid state NMR, exploiting
the fact of very long decoherence times in the rapidly fluctuating
liquid environment.\cite{NMR_Hore} Indeed, soon after Peter Shor
developed his prime factoring quantum algorithm, liquid state NMR
quantum computing efforts were the first out of the block.\cite{NMRqc}
In NMR quantum computing one employs a solvent hosting a concentration
of identical molecules with multiple nuclear spins (say protons).
Ideally, the chemical environment of the different nuclei are
different, so that they each have a different NMR chemical-shift, and
can thereby be addressed independently by varying the radio frequency.
In principle it is then possible to perform qubit operations on these
spins.  However, there are two major drawbacks to NMR quantum
computing -- the difficulty in scalability and the challenge of
preparing sufficiently entangled initial states.  As we now suggest,
it is possible that both of these can be circumvented by employing
small symmetric molecules which are the substrate for bond-breaking
enzymes.

By way of illustration, we consider the symmetric biochemical ion
pyrophosphate, P$_2$O$_7^{4-}$ (usually abbreviated as PPi), which is
important in metabolic activity.  Pyrophosphate is a phosphate dimer,
which consists of two phosphate ions, PO$_4^{3-}$, that share a
central oxygen.  (The inorganic-phosphate ion, abbreviated as Pi,
consists of a phosphorus atom tetrahedrally bonded to 4 oxygens.)
Since the phosphorus nuclei is a S$=1/2$ fermion and the oxygens are
$S=0$ bosons, the two-fold symmetry of PPi, which interchanges the two
$^{31}$P nuclei and the three end oxygens, will, like molecular
hydrogen, have two isomers, para-PPi and ortho-PPi.  Moreover, para-
and ortho-PPi will rotate with even and odd angular momentum,
respectively.  Thus ortho-PPi, with $L_\text{quasi}=1$, rotates with a
non-trivial Berry phase.

In biochemistry there is an enzyme (called pyrophosphatase) which
catalyzes the bond-breaking reaction, PPi $\rightarrow$ Pi + Pi.  Due
to the Berry phase term in ortho-PPi, we expect that this reaction
will be strongly suppressed, if not blocked entirely.  Then, provided
only para-PPi reacts, the two liberated Pi ions will have nuclear
spins which are entangled in a singlet.  Such intermolecular
entanglement of nuclear spins could, in principle, jump start liquid
state NMR quantum computing efforts, allowing for both scalability and
highly entangled initial state preparation.

\subsection{A new mechanism for isotope fractionation}

Isotope fractionation refers to processes that affect the relative
abundance of (usually) stable isotopes, often used in isotope
geochemistry and biochemistry.  There are several known mechanisms.
Kinetic isotope fractionation is a mass dependent mechanism in which
the diffusion constant of a molecule varies with the mass of the
isotope.  This process is relevant to oxygen evaporation from water,
where an oxygen molecule, which has one (or two) of the heavier oxygen
isotopes ($^{17}$O and $^{18}$O) is less likely to evaporate.  This
leads to a slight depletion in the isotope ratios of $^{17}$O/$^{16}$O
and $^{18}$O/$^{16}$O in the vapor relative to that in the liquid
water.

Another mass dependent isotope fractionation phenomena occurs in some
chemical reactions, where the isotope abundances in the products of
the reaction are (very) slightly different than in the reactants.  In
biochemistry this effect is usually ascribed to an isotopic
mass-induced change in the frequency of the molecular quantum
zero-point vibrational fluctuations when bonded in the pocket of an
enzyme.  This modifies slightly the energy of the activation barrier
which must be crossed in order for the bond-breaking reaction to
proceed.

However, there are known isotopic fractionation processes which are
``mass-independent'', a classic example being the increased abundance
of the heavier oxygen isotopes in the formation of ozone from two
oxygen molecules.\cite{fractionationMassIndep} In ozone isotope
fractionation the relative increased abundance of $^{17}$O and
$^{18}$O is largely the same.  While there have been theoretical
proposals to explain this ozone isotope anomaly, these are not without
controversy.\cite{fractionationExplain}

Here, as we briefly describe, our conjectured QDS rule for chemical
reactions involving small symmetric molecules leads naturally to the
prediction of a {\em new mechanism for isotope fractionation}, driven
by the quantum distinguishability of the two different isotopes.  In
the presence of isotopes that destroy the molecular rotational
symmetry, the QDS rule is no longer operative and one would expect the
chemical reaction to proceed more rapidly.

By way of illustration we again consider the enzymatic hydrolysis
reaction, PPi $\rightarrow$ Pi + Pi.  As we now detail, in this
experiment one would predict a large heavy oxygen isotope
fractionation effect.  Indeed, if one of the six ``end'' oxygens in
PPi is a heavy oxygen isotope the symmetry of PPi under a rotation is
broken and the reaction becomes ``unblocked'' (independent of the
nuclear spin state).

If correct, we would then predict a very large mass-independent oxygen
isotope fractionation which concentrates the heavy oxygen isotopes in
the products (Pi + Pi).  For the early stages of this reaction, before
the isotopically modified PPi are depleted, one would, in fact,
predict a factor of $4$ increase in the ratios of $^{17}$O/$^{16}$O
and $^{18}$O/$^{16}$O in the enzymatic reaction PPi $\rightarrow$ Pi +
Pi.

To be more quantitative we introduce a dimensionless function, $R(f)$,
where $R$ denotes the ratio of the heavy isotope of oxygen in the
products, relative to the reactants,
\begin{equation}
R(f) = \frac{ [^{18}\text{O}/ ^{16}\text{O} ]_\text{prod} } {[^{18}\text{O}/ ^{16}\text{O} ]_\text{react} } ,
\end{equation}
and $f\in [0,1]$ is the ``extent'' of the reaction.  In a conventional
isotope fractionation framework, one would expect a very small effect,
that is $R(f)\approx 1$.  But within our QDS conjecture, if correct,
one would have,
\begin{equation}
R(f) = \frac{ 1 - (1- f)^\lambda }{f} ,
\end{equation}
with $\lambda=4$.  Experiments to look for this effect are presently
underway.\cite{fractionationQDSexperiments}

\subsection{Activity of reactive oxygen species (ROS)}

In biochemistry it is well known that during ATP synthesis the oxygen
molecule picks up an electron and becomes a negatively charged
``superoxide'' ion, O$_2^{-}$.\cite{superOxide} Having an odd number
of electrons (with electron spin-$1/2$) superoxide is a ``free
radical''.  Together with hydrogen peroxide (H$_2 \text{O}_2$) and the
hydroxyl radical (the electrically neutral form of the hydroxide ion)
the superoxide ion is known as a Reactive Oxygen Species (ROS).  ROS
ions can cause oxidative damage due in part to their reactivity.
Indeed, in the free-radical theory, oxidative damage initiated by ROS
is a major contributor to aging. In biology there are specific enzymes
to break down the ROS to produce benign molecules
(e.g. water).\cite{ROSbio}

In contrast to the ROS, the stable state of molecular oxygen
(``triplet oxygen'') is less reactive in biology.  As we detail below,
we propose that this difference from triplet molecular oxygen can be
understood in terms of our conjectured QDS rule.

First we note that standard analysis of electronic molecular states
(that we relegate to the appendix) shows that under C$_2$ rotation the
electronic states in the triplet molecular (neutral) oxygen exhibit an
overall sign change. Because $^{16}$O nuclei are spinless bosons,
there is no nuclear contribution to the Berry phase.

Thus the triplet neutral oxygen molecule, O$_2$ exhibits a purely
electronic $\pi$ Berry phase, despite a spinless bosonic character of
the $^{16}$O nuclei. It therefore rotates with odd angular momentum,
$L=L_\text{quasi} + 2 {\bf Z}$, with $L_\text{quasi}=1$, identical to
ortho-hydrogen.  Our QDS conjecture then implies that {\it a direct
  bond-breaking chemical reaction of triplet oxygen is strictly
  forbidden.}

In contrast to triplet oxygen, the superoxide ion, O$_2^{-}$ is not
blocked by the QDS rule and can thus undergo a direct chemical
bond-breaking transition.  Indeed, as detailed in the appendix, due to
the electronic non-zero orbital and spin angular momenta aligned along
the body axis, the two ends of the superoxide ion are {\it
  distinguishable}.  Thus, in contrast to triplet oxygen, superoxide
does not have any symmetry under a 180 degree rotation that
interchanges the two oxygen nuclei.  The superoxide ion can thus
rotate with any integer value of the angular momentum,
$L=0,1,2,3,...$.

As a result, the QDS is not operative and thus there is no selection
rule precluding a direct bond-breaking chemical reaction of the
superoxide ion.  We propose that it is this feature of superoxide,
relative to triplet oxygen, which accounts, at least in part, for the
high reactivity of superoxide and explains why it is a ``Reactive
Oxygen Species''.

\subsection{Ortho-water as a Quantum Disentangled Liquid}

Molecular water has a C$_2$ symmetry axis which exchanges the two
protons.  Thus, as for molecular hydrogen, water comes in two
variants, para-water and ortho-water which rotate with even and odd
angular momentum, respectively.  QDS then predicts that the
ortho-water molecule (with $L_\text{quasi}=1$) can not undergo a
direct chemical reaction that splinters the molecule into a proton and
a hydroxide ion, OH$^{-}$.

Since the difference between the rotational kinetic energy of a
para-water and an ortho-water molecule is roughly 30K, liquid water
consists of 75\% ortho-water molecules and 25\% para-water molecules.
In one remarkable paper\cite{H20fractionation} it was reported that
gaseous water vapor can be substantially enriched in either ortho- or
para-water molecules, and then condensed to create ortho- and
para-liquid water -- although attempts to reproduce this work have
been unsuccessful.  If QDS is operative, ortho-liquid water would be
quite remarkable, having zero concentration of either ``free'' protons
or ``free'' hydroxide ions, despite these being energetically
accessible at finite temperature.  Theoretically, ortho-liquid water
would then be an example of a ``Quantum Disentangled Liquid'' in which
the protons are enslaved to the oxygen ions and do not contribute
independently to the entropy density.\cite{QDLpaper}

Experimentally, one would predict that ortho-liquid water would have
exactly zero electrical conductivity. Data on ``shocked''
super-critical water indicates that above a critical pressure the
electrical conductivity increases by 9 orders of
magnitude\cite{shockedH2O}.  Perhaps this is due to a transition from
a quantum-disentangled to a thermal state, where most (if not all) of
the ortho-water molecules are broken into a proton and hydroxide ion,
leading to a significant electrical conductivity?
 
The properties of ortho-solid ice might also be quite interesting,
provided QDS is operative.  While an extensive equilibrium entropy
would still be expected (consistent with the ice rules) the quantum
dynamics would be quite different.  Rather than protons hopping
between neighboring oxygen ions, in ortho-ice these processes would
actually correspond to collective rotations of the water molecules.
The nature of the quantum dynamical quenching of the entropy when ice
is cooled to very low temperatures is worthy of future investigation.
 
\section{Summary and conclusions}

In this paper we have explored the role of quantum
indistinguishability of nuclear degrees of freedom in enzymatic
chemical reactions.  Focussing on chemical bond-breaking in small
symmetric molecules, we argued that the symmetry properties of the
nuclear spins, which are entangled with -- and dictate -- the allowed
angular momentum of the molecules orbital dynamics, can have an order
one effect on the chemical reaction rate.  Our central thesis is a
Quantum Dynamical Selection(QDS) rule which posits that {\em direct}
bond-breaking reactions from orbitally asymmetric molecular states are
``blocked'', and only orbitally symmetric molecular states can undergo
a bond-breaking reaction.  This selection rule, which is not of
energetic origin, is quantum dynamical and for orbitally non-symmetric
molecular states involves the ``closing of a bottleneck'' in the
systems' Hilbert space.

The QDS rule is intimately linked to the importance of Fermi/Bose
indistinguishability of the nuclei during the enzymatic process which
implements a projective measurement onto orbitally symmetric molecular
states.  Mathematically, a Berry phase term, which encodes the
Fermi/Bose indistinguishability leads to an interference between the
multiple bond-breaking processes -- one for each of the symmetry
related molecular orbital configurations.  For an orbitally
non-symmetric molecular state this interference is destructive,
thereby closing off the bond-breaking reaction -- offering a
mathematical description of the QDS rule.
 
In much of this paper we focused on simple molecules with a planar
C$_n$ rotational symmetry about a particular molecular axis.  In this
case the Berry phase is determined by a ``quasi-angular-momentum'',
$L_\text{quasi}$, set by the symmetry of the nuclear spin
wavefunction.  For this planar case our QDS rule predicts that an
enzymatic bond-breaking transition implements a projective measurement
onto $L_\text{quasi} =0$.

Our QDS rule leads to a number of experimental implications that we
explored in Section \ref{sec:experiments}, including (i) a
differential chemical reactivity of para- and ortho-hydrogen, (ii) a
mechanism for inducing inter-molecular quantum entanglement of nuclear
spins, (iii) a new, mass-independent isotope fractionation mechanism,
(iv) a novel explanation of the enhanced chemical activity of
``Reactive Oxygen Species'' and (vi) illuminating the importance of
ortho-water molecules in modulating the quantum dynamics of liquid
water.  For each topic we have only scratched the surface, and a more
detailed exploration of these, and other experimental implications of
QDS, will be left for future work.
  
\appendix

\section{Electronic structure of the oxygen molecule and the
  ``superoxide'' ion, O$_2^{-}$}

Here we briefly review the electronic structure of both the stable
electrically neutral (triplet) oxygen molecule and of the superoxide
ion.  In each, we focus on the symmetry transformation properties of
the molecular electrons under a $\pi$ rotation.

First consider a single neutral oxygen atom, which has eight electrons
in the atomic orbitals, 1s$_2$2s$_2$2p$_4$.  It is convenient to
organize the three 2p-orbitals as p$_z$ and p$_\pm = \text{p}_x \pm i
\text{p}_y$, with wavefunctions in spherical coordinates,
$\psi_{\text{p}_z}= \cos(\theta)$ and $\psi_{\text{p}_\pm} = e^{\pm i
  \phi} \sin(\theta)$.

For two oxygen nuclei located at $\vec{R}_\pm = \pm a \hat{z}$, the
electronic molecular orbitals in order of increasing energy are the 1s
bonding ($\sigma_{1\text{s}}$), the 1s anti-bonding
($\sigma_{1\text{s}}^*$), the 2s bonding orbital
($\sigma_{2\text{s}}$), the 2s anti-bonding orbital
($\sigma_{2\text{s}}^*$), the $2\text{p}_z$ bonding orbital
($\sigma_z$), the two-fold degenerate $2\text{p}_\pm$ bonding orbitals
($\Pi_\pm$), the two-fold degenerate $2p_\pm$ antibonding orbitals
($\Pi^*_{\pm}$) and the $2p_z$ antibonding orbital ($\sigma_z^*$).
The molecular wavefunctions for the $2\text{p}_\pm$ antibonding
orbitals $\Pi^*_{\pm}$ are given by,
\begin{equation}
\Psi_{\Pi^*_\pm}({\bf r}) = e^{\pm i \phi(x,y)} \sum_{s=\pm1}  s \sin(\theta(z -s a))  .
\label{Pi_anti-bonding}
\end{equation}

The oxygen molecule has 16 electrons, 14 in spin singlet pairs in the
7 lowest energy molecular orbitals, ($\sigma_{1\text{s}},
\sigma_{1\text{s}}^*, \sigma_{2\text{s}}, \sigma_{2\text{s}}^*,
\sigma_z, \Pi_\pm$). We focus on the two electrons in the highest
occupied molecular orbitals (HOMO) $\Pi^*_{\pm}$.  Due to a molecular
Hund's rule, these two electrons are in a spin-aligned triplet state
with wavefunction,
\begin{equation}
\Psi_{\uparrow \uparrow}({\bf r}_1,{\bf r}_2) = \frac{1}{\sqrt{2}} [ \Psi_{\Pi_+^*}({\bf r}_1) \Psi_{\Pi_-^*}({\bf r}_2) - \Psi_{\Pi_+^*}({\bf r}_2) \Psi_{\Pi_-^*}({\bf r}_1)],
\end{equation}
which has been appropriately antisymmetrized $\Psi_{\uparrow
  \uparrow}({\bf r}_1,{\bf r}_2) = - \Psi_{\uparrow \uparrow}({\bf
  r}_2,{\bf r}_1)$.  From Eq.(\ref{Pi_anti-bonding}) this wavefunction
can be re-expressed as,
\begin{equation}
\Psi_{\uparrow \uparrow}({\bf r}_1,{\bf r}_2) = \sin(\phi_1 - \phi_2) 
\sum_{s_1,s_2 = \pm 1} s_1 s_2 \sin(\theta_1) \sin(\theta_2) ,
\end{equation}
with $\phi_j \equiv \phi(x_j,y_j)$ and $\theta_j =\theta(z_j - s_j a)$
for $j=1,2$.

We now consider performing a body rotation of the oxygen molecule by
180 degrees about an axis in the x-y plane with unit normal $\hat{n} =
\cos(\alpha) \hat{x} + \sin(\alpha) \hat{y}$ for some angle $\alpha$.
Under this transformation, $\phi \rightarrow 2 \alpha - \phi$ and
$\theta \rightarrow \pi - \theta$ (equivalently $z \rightarrow -z$).
Thus, one has, $\sin(\phi_1 - \phi_2) \rightarrow - \sin(\phi_1 -
\phi_2)$ and $\sin(\theta(z - s a)) \rightarrow \sin(\theta(z + sa))$,
implying that this 180 degree rotation induces a sign change in the
wavefunction for the two HOMO electrons,
\begin{equation}
\Psi_{\uparrow \uparrow}({\bf r}_1,{\bf r}_2) \rightarrow  - \Psi_{\uparrow \uparrow}({\bf r}_1,{\bf r}_2)  .
\end{equation}
Since the other molecular orbitals are occupied by two (spin-singlet)
electrons each contributes an overall plus sign, either (+1)$^2$ or
(-1)$^2$.  Thus, under C$_2$ rotation in triplet molecular oxygen
O$_2$ the electronic states exhibit an overall sign change.

In contrast, for superoxide the two HOMO orbitals are now occupied by
three electrons (two in one orbital, say $\Pi_+^*$, and the third in
$\Pi_-^*$).  These three electrons carry a net unit of angular
momentum aligned along the molecular axis of the oxygen molecule, $J_z
= \pm 1$ for the extra electron in the $\Pi_{\pm}^*$ molecular
orbital, respectively.  Together with the total electron spin which
take on two values, $S_z = \pm 1/2$, the molecule has a 4-fold
degeneracy $(J_z,S_z) = (\pm 1, \pm 1/2)$.  In the presence of
spin-orbit interactions, $H_\text{s-o} = - \lambda \vec{J} \cdot
\vec{S}$ this degeneracy is broken, and the molecular ground state is
a doublet with $(J_z,S_z) = (1, 1/2)$ and $(J_z,S_z) = (-1, -1/2)$, as
required by Kramer's theorem.
  
\acknowledgments 

We are deeply indebted and most grateful to Stuart Licht for general
discussions on this topic and especially for emphasizing the
importance of oxygen isotope fractionation experiments to access the
(putative) role of nuclear spins in the enzymatic hydrolysis of
pyrophosphate, which we discussed in Section V.C.  We would also like
to thank Jason Alicea, Leon Balents, Maissam Barkeshli, Steve Girvin,
Victor Gurarie, Andreas Ludwig, Lesik Motrunich, Michael Mulligan,
David Nesbitt, Nick Read, T. Senthil, Michael Swift, Ashvin Vishwanath
and Mike Zaletel for their patience and input on our work.

MPAF's research was supported in part by the National Science
Foundation under Grant No. DMR-14-04230, and by the Caltech Institute
of Quantum Information and Matter, an NSF Physics Frontiers Center
with support of the Gordon and Betty Moore Foundation.  MPAF is grateful to the Heising-Simons Foundation for support.  LR was
supported by the Simons Investigator Award from the Simons Foundation,
by the NSF under grant No. DMR-1001240, the NSF MRSEC Grant
DMR-1420736, and by the KITP under grant No.NSF PHY-1125915. LR thanks
the KITP for its hospitality as part of the Synthetic Matter workshop
and sabbatical program.


\end{document}